\documentclass[12pt,epsf]{article}
\usepackage{verbatim}
\usepackage{anyfontsize}
\usepackage{dsfont}
\usepackage{tikz}
\usepackage{sidecap}
\sidecaptionvpos{figure}{t}
\usepackage{subfigure}
\usepackage{enumitem}
\usepackage[english]{babel}
\usepackage{enumitem}  

\usepackage{amssymb,amsmath}
\usepackage{graphicx, xcolor, varwidth}
\usepackage{setspace}
\usepackage[permil]{overpic}
\usepackage{cite}
\usepackage{mathtools}

\DeclareSymbolFont{matha}{OML}{txmi}{m}{it}% txfonts
\DeclareMathSymbol{v}{\mathord}{matha}{118}

\usepackage[framemethod=default]{mdframed}
\newmdenv[skipabove=7pt,
skipbelow=7pt,
rightline=false,
leftline=false,
topline=false,
bottomline=false,
backgroundcolor=blue!10,
linecolor=blue,
innerleftmargin=5pt,
innerrightmargin=5pt,
innertopmargin=5pt,
innerbottommargin=5pt,
leftmargin=0cm,
rightmargin=0cm,
linewidth=4pt]{bBox}

\colorlet{darkblue}{blue!70!black}
\colorlet{darkgreen}{green!70!black}

\usepackage[colorlinks=true,urlcolor=darkblue,linktocpage=true,linkcolor=darkblue,citecolor=darkblue]{hyperref}
\hypersetup{ 
colorlinks=true, 
linkcolor=blue, 
citecolor=magenta, 
}

\numberwithin{equation}{section}
\DeclareMathSymbol{v}{\mathord}{matha}{118}
\newcommand{\be}{\begin{equation}}
\newcommand{\ee}{\end{equation}}
\newcommand{\bea}{\begin{eqnarray}}
\newcommand{\eea}{\end{eqnarray}}
\newcommand{\bear}{\begin{eqnarray}}
\newcommand{\eear}{\end{eqnarray}}
\newcommand{\beas}{\begin{eqnarray*}}
\newcommand{\p}{\partial}
\newcommand{\eeas}{\end{eqnarray*}}
\newcommand{\ba}{\begin{array}}
\newcommand{\ea}{\end{array}}

\def\ba#1\ea{\begin{align}#1\end{align}}
\def\bs#1\es{\begin{split}#1\end{split}}

\renewcommand{\r}{\rho}
\newcommand{\br}{\bar{\rho}}

\newcommand{\tr}{\operatorname{tr}}
\newcommand{\pd}[2][1]{\ifnum#1=1 \frac{\partial}{\partial {#2}} \else
  \frac{\partial^#1}{\partial {#2}^{#1}}\fi}
\newcommand{\dpd}[2][1]{\ifnum#1=1 \dfrac{\partial}{\partial {#2}} \else
  \frac{\partial^#1}{\partial {#2}^{#1}}\fi}
\newcommand{\td}[2][1]{\ifnum#1=1 \frac{d}{d{#2}} \else
  \frac{d^#1}{d{#2}^{#1}}\fi}

\renewcommand{\(}{\left(}
\renewcommand{\)}{\right)}

\renewcommand{\]}{\right]}

\newcommand{\nbox}{{\,\lower0.9pt\vbox{\hrule \hbox{\vrule height 0.2 cm \hskip 0.19 cm \vrule height 0.2 cm}\hrule}\,}}

\def\O{{\cal O}}

\newcommand{\N}{{\cal N}}



\begin{document}
\begin{spacing}{1.3}
\begin{titlepage}
  % end of \vbox

\begin{center}
{\Large 
\vspace*{6mm}

\bf 

Subleading Bounds on Chaos

}

\vspace*{6mm}

Sandipan Kundu

\vspace*{6mm}

\textit{Department of Physics and Astronomy,
\\ Johns Hopkins University,
Baltimore, Maryland 21218, USA\\}

\vspace{6mm}

{\tt \small kundu@jhu.edu}

\vspace*{6mm}
\end{center}

\begin{abstract}
Chaos, in quantum systems, can be diagnosed by certain out-of-time-order correlators (OTOCs) that obey the chaos bound of Maldacena, Shenker, and Stanford (MSS). We begin by deriving a dispersion relation for this class of OTOCs, implying that they must satisfy many more constraints beyond the MSS bound. Motivated by this observation, we perform a systematic analysis obtaining an infinite set of constraints on the OTOC. This infinite set includes the MSS bound as the leading constraint. In addition, it also contains subleading bounds that are highly constraining, especially when the MSS bound is saturated by the leading term. These new bounds, among other things, imply that the MSS bound cannot be exactly saturated over any duration of time, however short. Furthermore, we derive a sharp bound on the Lyapunov exponent $\lambda_2 \le \frac{6\pi}{\beta}$  of the subleading correction to maximal chaos.

\end{abstract}

\end{titlepage}
\end{spacing}

\vskip 1cm
\setcounter{tocdepth}{2}  
\tableofcontents

\begin{spacing}{1.3}

\newpage
%%%%%%%%%%%%%%%%%%%%%%%%%%%%%%%%%%%%%%%%%
\section{Introduction}
Quantum chaos is a remarkably complex phenomenon. Nevertheless, some theoretical understanding of chaos can be developed directly from basic principles. The important first step was to identify out-of-time-order correlators (OTOCs) as one of the fundamental probes of chaos in quantum systems \cite{1969JETP...28.1200L,Shenker:2013pqa,kitaev2014hidden,Maldacena:2015waa,Swingle:2018ekw}.

In recent years, a specific regularized thermal OTOC \cite{Maldacena:2015waa} 
\be\label{eq:otoc}
F(t)=\tr \[y V(0) y W(t)yV(0)yW(t)\]\ , \qquad y^4=\frac{e^{-\beta H}}{\tr\[e^{-\beta H}\]}\
\ee
of any two simple Hermitian local operators $V$ and $W$,\footnote{For simplicity, we assume that operators $V$ and $W$ have vanishing thermal one-point functions.} where temperature $T=1/\beta$,  has emerged as a good measure of quantum chaos since it has several appealing features. Physically, the OTOC (\ref{eq:otoc}) measures the effect of a small perturbation created by the operator $V$ on another operator $W$ at a later time $t>0$, hence determining the degree to which information of the initial perturbation becomes effectively lost. A technical advantage of the thermally regularized OTOC (\ref{eq:otoc}) is that  it is free from coincident point singularities in quantum field theory. Moreover, from a mathematical perspective, this OTOC enjoys certain analyticity properties, providing a way to utilize various tools from complex analysis. This observation led Maldacena, Shenker, and Stanford \cite{Maldacena:2015waa} to prove  a universal bound on the rate of growth of $F(t)$ in thermal quantum systems with a large number of degrees of freedom, revealing a profound insight into quantum chaos. However, there are compelling reasons to believe that the (MSS) chaos bound of \cite{Maldacena:2015waa} is part of a more general set of bounds. One of the goals of this paper is to find these additional bounds systematically.

The MSS bound essentially constrains the leading growing term of the early-time expansion of the regularized thermal OTOC (\ref{eq:otoc}). Since $F(t)$ is a bounded analytic function, we expect that  the subleading growing terms in the early-time expansion of $F(t)$ are also constrained. After all, not all early-time expansions can resum into well-behaved functions that are analytic and bounded even at late times. One significant hint in favor of additional bounds comes from conformal field theory (CFT).  It is well-known that certain CFT Regge correlators can be interpreted as a special case of the thermal OTOC (\ref{eq:otoc}) on Rindler space.\footnote{It is also known that the MSS bound is closely related to CFT causality constraints \cite{Hartman:2015lfa,Hartman:2016dxc,Perlmutter:2016pkf,Hofman:2016awc,Hartman:2016lgu,Afkhami-Jeddi:2016ntf,Afkhami-Jeddi:2017idc,Afkhami-Jeddi:2017rmx,Afkhami-Jeddi:2018own,Chowdhury:2018uyv}. Moreover, there is also a connection between the MSS bound and the 4D $a$-theorem \cite{Kundu:2020bdn}.} Recently, studies of these CFT Regge correlators have yielded an infinite set of constraints \cite{Kundu:2020gkz,Kundu:2021qpi} beyond the MSS bound. So, it is a natural question to ask whether similar chaos bounds exist in general quantum systems with many degrees of freedom.

In this paper, we answer this question starting from the basic properties of the OTOC (\ref{eq:otoc}). This OTOC, under very general assumptions about physical systems \cite{Maldacena:2015waa}, has the following properties: in the half-strip $\{t\in \mathbb{C}|\ \mbox{Re}\ t\ge t_0\ \text{and}\ |\mbox{Im}\ t|\le \frac{\beta}{4} \}$, $F(t)$  is (i) analytic, and  (ii) bounded by the factorized correlator $|F(t)|\le F_d$ on the boundary $\ |\mbox{Im}\ t|= \frac{\beta}{4},\ \mbox{Re}\ t\ge t_0$,  (iii) obeying the Schwarz reflection condition $\(F(t)\)^*=F\(t^*\)$. The {\it factorization time} $t_0$ will  play an important role throughout. Specifically,  it is the time scale after which the time ordered thermal correlator factorizes $\tr \[y^2 W(t) V(0) y^2 V(0)W(t)\]\approx F_d$.\footnote{The factorized correlator is defined in equation (\ref{def:Fd}).} The above conditions are expected to hold for all unitary quantum systems with a large number of degrees of freedom and a Hamiltonian $H$ which is a finite product of simple operators. Moreover, for such systems, it is also expected that there is a parametric separation between  the factorization time scale $t_0$ and the {\it scrambling time} $t_*$.\footnote{The scrambling time is the time scale at which $F(t)$ starts decreasing rapidly \cite{Sekino:2008he}.} The MSS bound  \cite{Maldacena:2015waa} is a local consistency condition 
\be\label{intro:MSS}
 \( \frac{2\pi}{\beta}- \p_t\)\(F_d-F(t)\)\ge 0\ ,  \qquad \qquad   t\gg  t_0
\ee
which follows directly from properties (i)-(iii) of the OTOC. It should be noted that originally a stronger version\footnote{The stronger boundedness condition of \cite{Maldacena:2015waa} additionally requires that $|F(t)|\le F_d$ for $\mbox{Re}\ t=t_0$ and $| \mbox{Im}\ t|\le \beta/4$. So, it was assumed in \cite{Maldacena:2015waa} that the factorization time $t_0$ can always be chosen such that the OTOC obeys this additional condition. In this paper, we do not make this additional assumption and hence  our $t_0$ is conceptually very similar to the dissipation time scale. } of the boundedness condition was used in \cite{Maldacena:2015waa} to derive (\ref{intro:MSS}). However, we will show that the above weaker boundedness condition is sufficient  for the derivation of the MSS bound and its generalizations.

In chaotic systems, the OTOC starts decreasing rapidly for $t\gg t_0$, only after the onset of scrambling. In particular, a signature of strong chaos is the exponential growth \cite{Liao:2018uxa,Lewis-Swan:2019pln}  of the leading order term $F_d-F(t)\sim e^{\lambda_L t}$ where  $\lambda_L$ is the Lyapunov exponent. So, the MSS bound (\ref{intro:MSS}) places a universal upper bound on the Lyapunov exponent $\lambda_L\le \frac{2\pi}{\beta}$. Importantly, all large $N$ theories that are holographically dual to Einstein gravity saturate the MSS bound at the leading order in $1/N$ \cite{Roberts:2014isa,Shenker:2013pqa,Shenker:2013yza,kitaev2014hidden,Shenker:2014cwa}.

The MSS bound does not fully utilize the analyticity, boundedness, and Schwarz reflection properties of $F(t)$. In fact the MSS bound can be rederived as part of an infinite set of constraints that can be organized systematically  by defining a local {\it moment of the OTOC} 
\be\label{def:moments}
\mu_J\(t\)=e^{\frac{4\pi J}{\beta}t} \int_{t-i \frac{\beta}{4}}^{t+i \frac{\beta}{4}} dt' e^{-\frac{2\pi  }{\beta}\(t'-i \frac{\beta}{4}\)(2J+1)}  \(F(t')-F_d\)
\ee 
for real $t\ge t_0$ and any $J$. Note that for integer $J\ge 0$, moments $\mu_J(t)$ are real. The moments $\mu_J(t)$ provide an alternative but equivalent description of quantum chaos. We will show that analyticity, boundedness, and Schwarz reflection properties of the OTOC imply the following local consistency conditions for $\mu_J(t)$ for all integer $J\ge 0$ and $t\ge t_0$:
\begin{itemize}
\item{{\bf Positivity \& Boundedness}--
\be\label{intro:positive}
0<\mu_J(t)<\frac{2\beta F_d}{\pi(2J+1)}e^{-\frac{2\pi}{\beta}t}\ ,
\ee}
\item{{\bf Monotonicity}--
\be\label{intro:monotonic}
\mu_{J+1}(t)<\mu_J(t)\ ,
\ee
}
\item{{\bf Log-Convexity}--
\be\label{intro:log}
\mu_{J+1}(t)^2\le \mu_J\(t\)\mu_{J+2}\(t\)\ .
\ee
}
\end{itemize}
Furthermore, the rate of growth of $\mu_J(t)$, for integer $J$,  is also bounded  for $t\ge t_0$. The above constraints are the main results of this paper.

Let us now consider a chaotic system that exhibits a period of exponential growth $F_d-F(t)\propto \exp(\lambda_L t)$ over some duration of time. Over the same duration of time, one can easily calculate the moments (\ref{def:moments}) for any $J$. The MSS bound on the Lyapunov exponent $\lambda_L\le \frac{2\pi}{\beta}$ follows directly from the positivity condition (\ref{intro:positive}). Besides, consistency conditions (\ref{intro:positive})-(\ref{intro:log}) place further constraints on growing terms that are subleading, as we discuss in section \ref{sec:implications}. These additional constraints are particularly important when the MSS bound is saturated.\footnote{The chaos bound (\ref{intro:MSS}), as shown in \cite{Murthy:2019fgs}, also follows naturally from the eigenstate thermalization hypothesis \cite{PhysRevA.43.2046,PhysRevE.50.888,Srednicki_1999,Rigol_2008,D_Alessio_2016}. It would be interesting to explore whether this connection exists even for the subleading chaos bounds. }  

A period of maximal chaos can be equivalently described by the associated moments $\mu_0(t)=$ constant and $\mu_J(t)=0$ for all integer $J\ge 1$. Clearly, this is inconsistent with the strict positivity (\ref{intro:positive}) and the monotonicity (\ref{intro:monotonic}) conditions implying that the term $F_d-F(t)\sim \exp(\frac{2\pi}{\beta} t)$ alone, in any time duration, violates these new chaos bounds. Hence, subleading corrections to this leading Lyapunov growth are necessarily required long before the scrambling time $t_*$.  We parametrize the subleading corrections as follows
\be\label{intro:para}
F_d - F(t)= \frac{1}{\N}\(c_1e^{\frac{2\pi}{\beta} t}+ c_2 \varepsilon e^{\lambda_2 t}+\cdots\) \qquad \qquad   t_*\gg t\gg t_0
\ee
where, $c_1,c_2$ are order one coefficients and $\N\gg 1$ is a measure of the effective number of degrees of freedom per site determining the scrambling time $t_*=\frac{1}{2\pi}\ln \N$ \cite{Sekino:2008he}. Whereas,  $\varepsilon>0$ is a small dimensionless parameter\footnote{Note that  $\varepsilon$ can just be additional powers of $\frac{1}{\N}$. However, in general, it can also be a completely independent parameter. In holographic theories, $\frac{1}{\N}$ has the interpretation of the Newton constant $G_N$. The parameter $\varepsilon$, when independent, can be thought of as the analog of the string coupling. Of course, the bound (\ref{bound:intro}) applies to both these scenarios.} ensuring that the second term on the right-hand side is subleading. We will argue that there is also a universal upper bound on $\lambda_2$:
\be\label{bound:intro}
\lambda_2 \le \frac{6\pi}{\beta}
\ee
that follows from the constraints (\ref{intro:positive})-(\ref{intro:log}). At first sight, there appears to be a tension between the above bound and the MSS bound since the subleading term in (\ref{intro:para}) can grow faster than $ \exp(\frac{2\pi}{\beta} t)$. However, this is not actually a problem. The same set of constraints (\ref{intro:positive})-(\ref{intro:log}) also implies that the approximation (\ref{intro:para}) must break down before the subleading growing term becomes the dominant contribution.

Likewise, if the subleading growing term also saturates the bound (\ref{bound:intro}), then there is a universal bound on the next order term of (\ref{intro:para}). In particular, the Lyapunov exponent of the next order term, when $\lambda_2 = \frac{6\pi}{\beta}$, must obey $\lambda_3 \le \frac{10\pi}{\beta}$. We argue that the same pattern persists even for higher-order correction terms. Furthermore, the coefficients of these higher-order growing terms are also highly constrained, as we discuss in section  \ref{sec:implications}.

There are some maximally chaotic systems where subleading terms of  $F(t)$ can also be studied. For example, in the Schwarzian theory, it is possible to compute the OTOC (\ref{eq:otoc}) as an expansion in $\beta/C = 1/\N$ \cite{Maldacena:2016upp,Lam:2018pvp}. The subleading term in that expansion does obey the bound (\ref{bound:intro}). The same is true even in 2d CFT, as one can readily see from \cite{Roberts:2014ifa}. Moreover, the bound (\ref{bound:intro}) is also consistent with our expectations from the Sachdev-Ye-Kitaev (SYK) model \cite{Stanford:2015owe,Gu:2018jsv,Kobrin:2020xms}. Interestingly, in all these examples $\lambda_2 = \frac{4\pi}{\beta}$, suggesting that there could be a stronger bound on $\lambda_2$. However, any such stronger bound will definitely require additional assumptions that are not true in general. After all, there are known examples that saturate the bound (\ref{bound:intro}). As we mentioned before, CFT Regge correlators are a special case of the OTOC (\ref{eq:otoc}). These Regge correlators saturate the bound (\ref{bound:intro}) in large-$N$ CFTs that are dual to effective field theories in AdS with higher derivative four-point interactions \cite{JP,Chandorkar:2021viw,Kundu:2021qpi}. We will discuss this in section \ref{sec:CFT}.

The rest of the paper is organized as follows. We begin with a review of the OTOC (\ref{eq:otoc}) and the MSS chaos bound in section \ref{sec:otoc}. In this section, we also derive a dispersion relation to demonstrate that the OTOC satisfies additional consistency conditions beyond the MSS bound.  In section \ref{sec:proof}, first we introduce the moments of the OTOC and then derive the constraints (\ref{intro:positive})-(\ref{intro:log}). In section \ref{sec:implications} we discuss some implications of these general constraints. In section \ref{sec:CFT} we show that the chaos bounds obtained in this paper are completely consistent with CFT bounds of \cite{Kundu:2020gkz,Kundu:2021qpi}. Finally, we conclude with some remarks in section \ref{sec:conclusions}.

\section{Chaos and OTOC}\label{sec:otoc}
In thermal quantum systems, there are three time scales that are relevant for chaos. A thermal two-point function $\tr\[y^4 V(0)V(t)\]$ of any simple Hermitian local operator $V$  decays exponentially $\sim e^{-t/t_d}$, where the decay time $t_d$ is generally known as the {\it dissipation time}.

In this paper, we mainly consider quantum systems with a large number of degrees of freedom. In such systems, the OTOC $F(t)\approx F_d$ before the onset of chaos, where $F_d$ is the factorized correlator 
\be\label{def:Fd}
F_d= \tr\[y^2 V(0) y^2 V(0)\]\tr\[y^2 W(t)y^2W(t)\]
\ee
which is time independent because of time translation invariance. A clear signature of quantum chaos is the deviation of $F(t)$ from the factorized value for $t\gg t_d$, irrespective of the choice of operators.  The {\it scrambling time} $t_*$ is the time scale at which $F(t)$ starts decreasing rapidly, such that $F_d-F(t_*) \sim \O(1) F_d$. 
For systems with a large number of degrees of freedom, it is expected that $t_d\sim \beta$ and a parametric separation between $t_d$ and $t_*$, provided  the Hamiltonian is sufficiently local containing finite products of simple operators  \cite{Maldacena:2015waa}.  

\subsection{Assumptions}\label{sec:assumptions}
\begin{figure}
\centering
\includegraphics[scale=0.5]{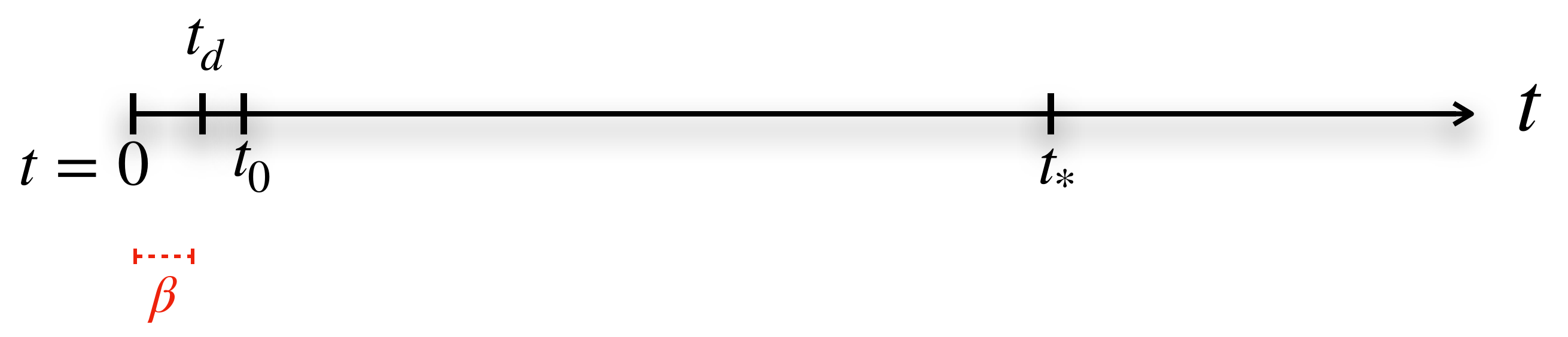}
\caption{ \label{fig:timescales} \small Three time scales that are important for chaos are shown here schematically for thermal quantum systems with a large number of degrees of freedom and a simple Hamiltonian which is sufficiently local. These time scales are: $t_d$ = dissipation time $\sim \beta$, $t_0$ = factorization time, and $t_*=$ scrambling time.}
\end{figure}

The MSS bound \cite{Maldacena:2015waa} can be derived directly from the following assumptions about the quantum system:
\begin{itemize}
\item[(1)] {\bf Unitarity}: The quantum system is unitary. 
\item[(2)] {\bf Factorization}: Time ordered correlators factorize at late times $t> t_d$. More precisely, we assume that for any  $V$ and $W$ these is some {\it factorization time} $t_0$ such that to a good approximation 
\be\label{eq:factorization}
\tr \[y^2 W(t) V(0) y^2 V(0)W(t)\]= F_d \qquad \text{for}\qquad t\ge t_0\ .
\ee
Of course, $t_0$ is never much different from the dissipation time scale, however, in a specific quantum system it can depend on the choice of operators. 
\item[(3)] {\bf Separation of scales}: The scrambling time is much larger than the other time scales: $t_* \gg t_0, t_d$.   
\end{itemize}
The last two assumptions, which are not really independent, are expected to be valid for a wide class of thermal quantum systems, especially for systems with a large number of degrees of freedom and a simple Hamiltonian which is sufficiently local. For such systems
\be\label{eq:factorization_corr}
F(t)= F_d+\frac{1}{\N}\O(1) \qquad \text{for}\qquad t\sim t_0\ ,
\ee
where $\N\gg 1$ is a measure of the number of degrees of freedom per site. A signature of strong chaos is the Lyapunov behavior 
\be\label{Lyapunov}
F_d-F(t)=\frac{c_1}{\N} e^{\lambda_L t}+\cdots \ , \qquad \text{for}\qquad t_d,t_0\ll t\ll t_*
\ee
where $c_1$ is a positive order one constant and $\lambda_L$ is the Lyapunov exponent. The OTOC $F(t)$ deviates significantly from the factorized value $F_d$ for $t\sim t_*$ and hence the scrambling time scales as $\ln \N$. Various time scales that are relevant for chaos are shown schematically in figure \ref{fig:timescales}.

It should be noted that the above factorization assumption is weaker than the factorization assumption of \cite{Maldacena:2015waa}. In \cite{Maldacena:2015waa}, it was also assumed that $|F(t)|\le F_d$ for $\mbox{Re}\ t=t_0$ and $| \mbox{Im}\ t|\le \beta/4$. However, this additional requirement is not actually necessary for the derivation of the MSS bound. We will establish this fact at the end of this section by deriving a dispersion relation for the OTOC.  

\subsection{Analyticity and Positivity Conditions}
Under the above assumptions, the OTOC $F(t)$ as defined in (\ref{eq:otoc}), has the following properties:
\begin{itemize}
\item[(i)] {\bf Analyticity}: $F(t)$, as a function of complex $t$, is analytic in the half strip \cite{Maldacena:2015waa}: 
\be\label{halfstrip}
\mbox{Re}\ t >0 \qquad \text{and} \qquad -\frac{\beta}{4}\le \mbox{Im}\ t \le \frac{\beta}{4}\ ,
\ee
as shown in figure \ref{figure:sumrule}. 
\item[(ii)] {\bf Boundedness}: $F_d$, as defined in (\ref{def:Fd}), is positive definite. Moreover, it was also shown in \cite{Maldacena:2015waa} that on the boundary of the half strip (\ref{halfstrip})
\be\label{eq:positive}
|F(t\pm i \beta/4)|\le F_d \qquad  \text{for} \qquad  t\ge t_0 \ .
\ee  
This follows from the factorization condition (\ref{eq:factorization}) and unitarity.\footnote{In \cite{Maldacena:2015waa}, a stronger version of the boundedness condition 
\be
|F(t)|\le F_d \qquad  \text{for} \qquad  \mbox{Re}\ t\ge t_0 \qquad  \text{and} \qquad  |\mbox{Im}\ t| \le \beta/4\nonumber
\ee
was utilized to derive the MSS bound. This stronger condition additionally requires that $|F(t)|\le F_d$ for $\mbox{Re}\ t=t_0$ and $| \mbox{Im}\ t|\le \beta/4$. This additional  assumption is not actually necessary. The weaker condition (\ref{eq:positive}) is sufficient for the derivation of the MSS bound and its generalizations.} 
\item[(iii)] {\bf Schwarz Reflection}: $F(t)$, as a function of complex $t$, satisfies the Schwarz reflection condition 
\be\label{eq:schwarz}
\(F(t)\)^*=F\(t^*\)
\ee
implying $F(t)$ is real for $\mbox{Im}\ t=0$. This follows from the fact that  $V$ and $W$ are local Hermitian operators.
\end{itemize}
Of course, the OTOC $F(t)$ is also periodic $F(t+i \beta)=F(t)$. 

\subsection{Maldacena-Shenker-Stanford Bound}
The MSS bound \cite{Maldacena:2015waa} is a constraint on the rate of growth of $F_d-F(t)$. In particular, the bound states  
\be\label{bound:MSS}
\frac{d}{dt}\(F_d-F(t)\) \le \frac{2\pi}{\beta}\(F_d-F(t)\)  \qquad \text{for}\qquad t\gg  t_0\ ,
\ee
where $t_0$ is the factorization time which was defined before. 

The above condition leads to a universal bound on the Lyapunov exponent for systems with $t_*\gg t_d, t_0$. For such systems, we can use the form (\ref{Lyapunov}) to obtain \cite{Maldacena:2015waa}
\be
\lambda_L \le \frac{2\pi}{\beta}\ .
\ee
In this paper, we will rederive the MSS chaos bound as a special case of more general constraints that follow  directly from the properties (\ref{halfstrip})-(\ref{eq:schwarz}) of the OTOC.

\subsection{Dispersion Relation and Consistency Conditions}
\begin{figure}
\centering
\includegraphics[scale=0.45]{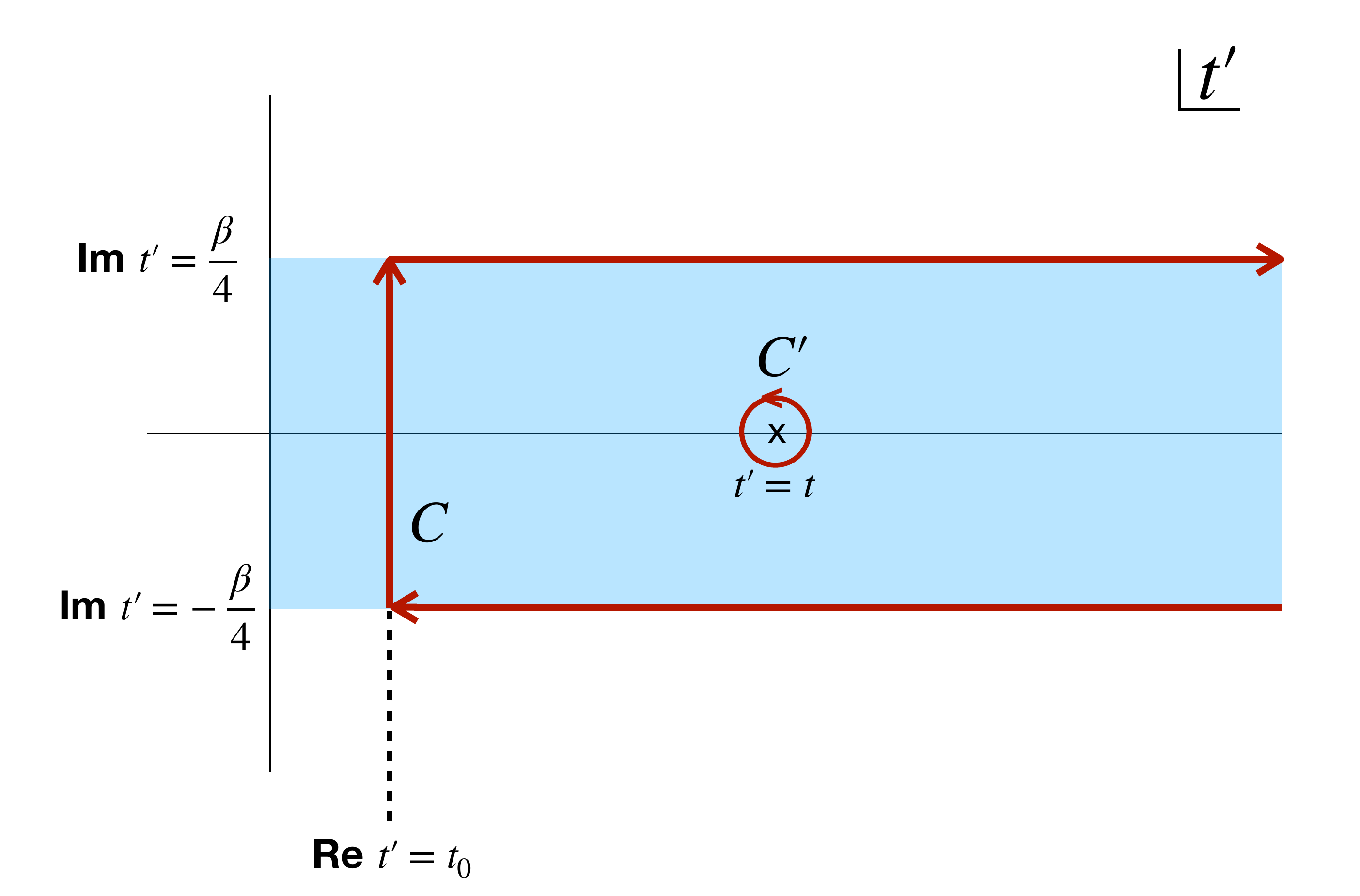}
\caption{ \label{figure:sumrule} \small The OTOC $F(t)$, as defined in (\ref{eq:otoc}), is analytic in the shaded blue region. Contours for deriving the dispersion relation (\ref{OTOC:sumrule}), where $t_0$ is the factorization time, are shown in red. The contour $C'$ surrounds a simple pole at $t'=t$ of the integral (\ref{eq:start}).}
\end{figure}

Let us now utilize the above properties (\ref{halfstrip})-(\ref{eq:schwarz}) of the OTOC $F(t)$ to derive a set of consistency conditions that generalize (\ref{bound:MSS}). We begin by using the analyticity condition to write
\be\label{eq:start}
\oint_C dt' \ e^{\frac{2\pi }{\beta}(t'-t)} \frac{F(t')-F_d}{1-e^{\frac{4\pi }{\beta}(t'-t)}}=-\oint_{C'} dt' \ e^{\frac{2\pi }{\beta}(t'-t)} \frac{F(t')-F_d}{1-e^{\frac{4\pi }{\beta}(t'-t)}}=\frac{i \beta}{2}  (F(t)-F_d)
\ee
where the contour $C$, as shown in figure \ref{figure:sumrule}, is a rectangle around the point $t$ with four sides at $\mbox{Re}\ t'=\mbox{Re}\ t_0$, $\mbox{Re}\ t'=\infty$, and $\mbox{Im}\ t'=\pm \beta/4$. We can ignore the contribution from the contour at infinity since $|F(t)|$ is bounded in the half strip (\ref{halfstrip}) for $\mbox{Re}\ t \ge t_0$. Hence, the above equation simplifies
\begin{align}\label{eq:middle}
F_d-F(t)=\frac{2}{\beta} e^{-\frac{2\pi (t-t_0)}{\beta}}& \int^{\frac{\beta}{4}}_{-\frac{\beta}{4}} d\tau \frac{e^{\frac{2\pi i \tau}{\beta}}\(F_d-F(t_0+i \tau)\)}{1- e^{\frac{4\pi i \tau}{\beta}}e^{-\frac{4\pi (t-t_0)}{\beta}}}\nonumber\\
&+\frac{2}{\beta} \int_{t_0}^\infty dt' \frac{\(F_d-\mbox{Re}\ F(t'+i\beta/4)\)}{\text{cosh}\(\frac{2\pi}{\beta}(t-t')\)}
\end{align}
for $|\mbox{Im}\ t |< \frac{\beta}{4}$. We now consider the limit $\mbox{Re}\ t\gg t_0$. In this limit, the first term in the above equation decays exponentially.\footnote{Note that in the limit $\mbox{Re}\ t\gg t_0$ the integral in the first term of equation (\ref{eq:middle}) is $ \O(1)/\N$ because $F(t_0)$ factorizes up to $1/\N$ correction terms. Hence, the first term of equation (\ref{eq:middle}) decays exponentially and can be safely ignored for $\mbox{Re}\ t\gg t_0$. }  This enables us to write a dispersion relation  
\be\label{OTOC:sumrule}
F_d-F(t)|_{{\rm Re}\ t\gg t_0}=\frac{2}{\beta} \int_{t_0}^\infty dt' \frac{\(F_d-\mbox{Re}\ F(t'+i\beta/4)\)}{\text{cosh}\(\frac{2\pi}{\beta}(t-t')\)} +\O\(e^{-\frac{2\pi (t-t_0)}{\beta}}\)\ .
\ee
All OTOCs, under the assumptions made in section \ref{sec:assumptions}, satisfy this dispersion relation\footnote{The OTOC dispersion relation (\ref{OTOC:sumrule}) is very similar to the CFT dispersion relations of \cite{Bissi:2019kkx,Carmi:2019cub}.} for $|\mbox{Im}\ t |< \frac{\beta}{4}$. Note that the condition (\ref{eq:positive}) implies that the integrand in (\ref{OTOC:sumrule}) is positive for real $t$.

The above dispersion relation leads to consistency conditions that all OTOCs must obey. These conditions can be obtained by constructing operators $D(t,\p_t)$ that satisfy 
\begin{align}
D(t,\p_t)\text{sech}\(\frac{2\pi}{\beta}(t-t')\) \ge 0 \qquad \text{for all} \qquad  t,t'\ge t_0 \ ,\label{eq:ccD}
\end{align}
where $t,t'$ are real. For any such operator, the dispersion relation (\ref{OTOC:sumrule}) immediately implies 
\be
D(t,\p_t)\(F_d-F(t)\)\ge 0 \qquad \text{for} \qquad  t\gg  t_0\ . \label{eq:cc2}
\ee
For example, one can check that the operator $D= \frac{2\pi}{\beta}- \p_t$ satisfies (\ref{eq:ccD}) establishing (\ref{bound:MSS}). Hence, the MSS bound is contained in the dispersion relation (\ref{OTOC:sumrule}). This also implies that the MSS bound follows from the weaker set of conditions (\ref{halfstrip})-(\ref{eq:schwarz}), as mentioned before.

The important point that we want to emphasize is that consistency conditions (\ref{eq:cc2}) are, in general, stronger than the MSS bound.  In order to demonstrate that we construct a set of operators that satisfy (\ref{eq:ccD}), leading to the following generalization of the MSS bound:
\be\label{eq:cc3}
\left[\prod_{I=1}^N \( \frac{2\pi(2I-1)}{\beta}- \p_t\)\right]\(F_d-F(t)\)\ge 0\ ,  \qquad  t\gg  t_0
\ee
for all integer $N\ge 1$. 

These bounds are particularly useful when the MSS bound is saturated. For example, consider the OTOC (\ref{intro:para}). The bound (\ref{eq:cc3}), for $N=1$ and $2$, then implies (\ref{bound:intro}).

It is needless to say that the consistency conditions (\ref{eq:cc2}) depend heavily on the starting integral (\ref{eq:start}). Moreover, it is also unclear how to pick a set of operators $D(t,\p_t)$ that lead to optimal bounds. So, at this point the goal would be to derive bounds more systematically by fully utilizing the properties (\ref{halfstrip})-(\ref{eq:schwarz}). This will be achieved by introducing local moments of the OTOC, as we describe next.

\section{Subleading Chaos Bounds}\label{sec:proof}
\begin{figure}
\centering
\includegraphics[scale=0.45]{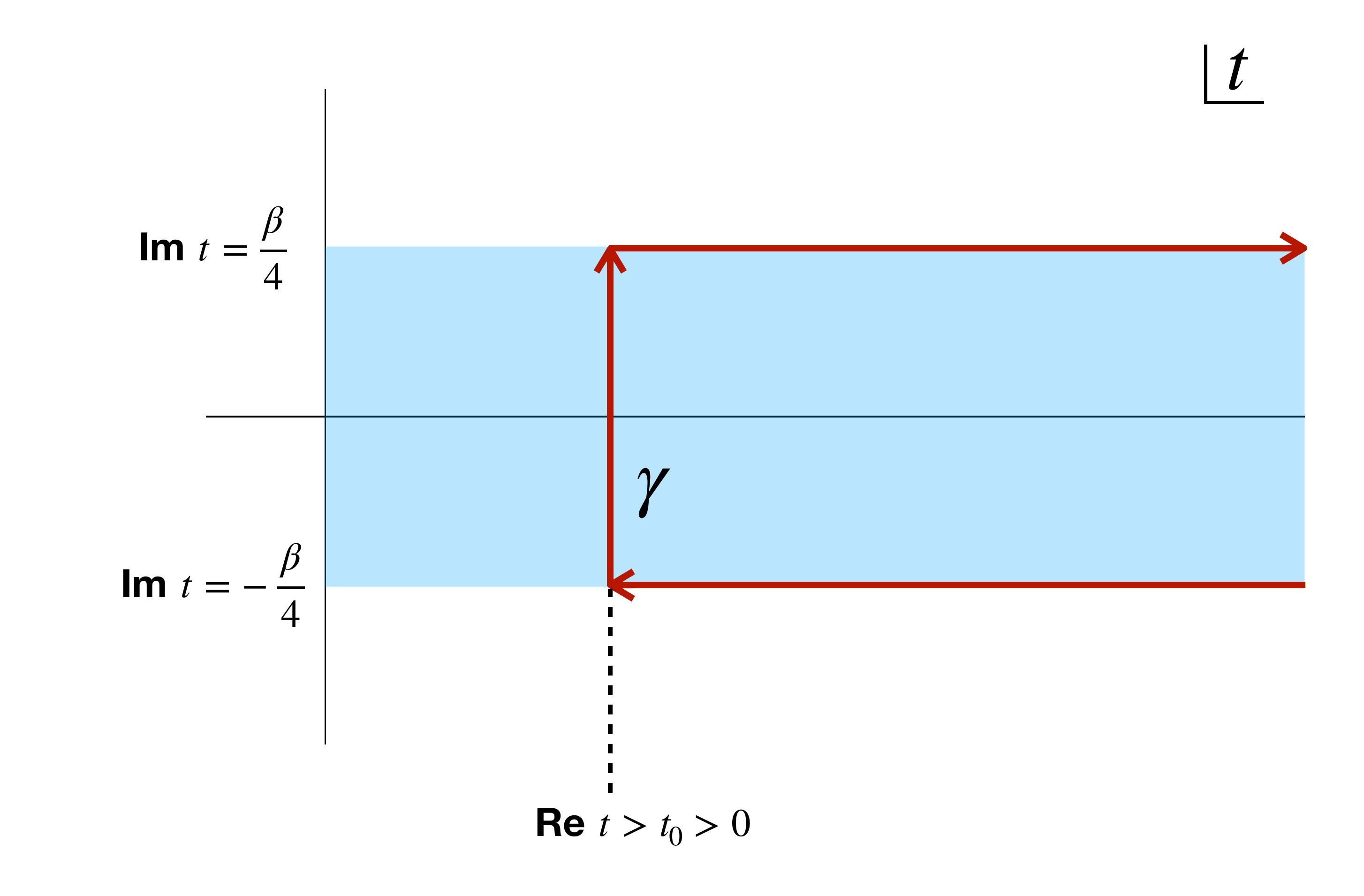}
\caption{ \label{contour} \small The contour $\gamma$ is used for deriving bounds on moments of the OTOC. }
\end{figure}

\subsection{Moments of the OTOC}
We begin by using the analyticity condition of $F(t)$ to write
\be
\oint_\gamma dt' P(t') \(F_d-F(t')\)=0\ ,
\ee
where the contour $\gamma$ is shown in figure \ref{contour} and $P(t)$ is an analytic function in the half-strip (\ref{halfstrip}) with the following properties: (i) $P(\mbox{Re}\ t\rightarrow \infty)\rightarrow 0$, (ii) $P(\mbox{Im}\ t =\beta/4)>0 $, and (iii) $P(\mbox{Im}\ t =-\beta/4)=- P(\mbox{Im}\ t =\beta/4)$.  The above contour integral and the Schwarz reflection condition (\ref{eq:schwarz}) then imply 
\be\label{eq:odd}
- \int_{t-i \frac{\beta}{4}}^{t+i \frac{\beta}{4}} dt' P(t') \(F_d-F(t')\)= 2\int_{t}^\infty dt' P\(t'+i\beta/4\)\(F_d-\mbox{Re}\ F(t'+i\beta/4)\)\ge 0
\ee
for any $t\ge t_0$. We have ignored the contribution from the contour at infinity since $P(\mbox{Re}\ t\rightarrow \infty)\rightarrow 0$. Of course, the positivity follows from the property (\ref{eq:positive}).

The set of functions $ \exp(-\frac{2\pi  }{\beta}\(t-i \frac{\beta}{4}\)(2J+1))$, for integer $J\ge 0$, form a basis for $P(t)$ functions. This naturally leads to the following definition for a moment of the OTOC for real $t$:
\be\label{def:moment0}
\mu_0\(t\)= \int_{t-i \frac{\beta}{4}}^{t+i \frac{\beta}{4}} dt' e^{-\frac{2\pi  }{\beta}\(t'-i \frac{\beta}{4}\)}  \(F(t')-F_d\)
\ee 
which we will refer to as the {\it primary moment} of the OTOC. We will show that the late-time ($t\gg \beta$) behavior of $F(t)$ is completely fixed by this primary moment. 

Similarly, we define higher moments $\mu_J\(t\)$ of the OTOC by (\ref{def:moments}) for any $J$ and real $t\ge t_0$. The prefactor $e^{\frac{4\pi J}{\beta}t}$ in (\ref{def:moments}) has been included for later convenience. Notice that moments (\ref{def:moments}) are well-defined even for negative and/or non-integer $J$. However, equation (\ref{eq:odd}) applies to $\mu_J\(t\)$ only when $J$ is a non-negative integer.

\subsection{Bounds on Moments}\label{sec:BoundsOnMoments}
We now derive a set of local consistency conditions of $\mu_J(t)$, for all integer $J\ge 0$,  directly from analyticity, boundedness, and Schwarz reflection properties of the OTOC.
\subsubsection{Positivity Condition }
When $J\ge 0$ is an integer, we can express $\mu_J(t)$ as an integral of a positive function by using equation  (\ref{eq:odd}) 
\be\label{eq:dispersion}
\mu_J\(t\)= 2e^{\frac{4\pi J}{\beta}t}\int_{t}^\infty dt' e^{-\frac{2\pi  (2J+1)}{\beta}t'}\(F_d-\mbox{Re}\ F(t'+i\beta/4)\)> 0 
\ee
for any real $t\ge t_0$. The strict positivity follows from the fact that the saturation of the above inequality necessarily requires $F(t)=F_d$  on the entire half-strip (\ref{halfstrip}). Hence, all chaotic systems, under the assumptions made in the previous section, must obey the strict positivity condition (\ref{eq:dispersion}). Note that a similar positivity condition does not exist when $J$ is negative.

\subsubsection{Boundedness Condition}
The equality in (\ref{eq:dispersion}), along with the boundedness property (\ref{eq:positive}), also leads to an upper bound 
\begin{align}\label{eq:bounded}
\mu_J\(t\)&= 2e^{\frac{4\pi J}{\beta}t}\int_{t}^\infty dt' e^{-\frac{2\pi  (2J+1)}{\beta}t'}\(F_d-\mbox{Re}\ F(t'+i\beta/4)\)\nonumber\\
&< 4e^{\frac{4\pi J}{\beta}t}F_d \int_{t}^\infty dt' e^{-\frac{2\pi  (2J+1)}{\beta}t'}=\frac{2\beta F_d}{\pi(2J+1)}e^{-\frac{2\pi }{\beta}t}
\end{align}
for any real $t\ge t_0$ and integer $J\ge 0$.

\subsubsection{Monotonicity Condition}
We observe that for integer $J\ge 0$ and real $t\ge t_0$
\begin{align}\label{eq:monotonicity}
\frac{\mu_{J+1}\(t\)}{\mu_J\(t\)}=\frac{\int_{t}^\infty dt' e^{-\frac{4\pi  }{\beta}(t'-t)} e^{-\frac{2\pi  (2J+1)}{\beta}t'}\(F_d-\mbox{Re}\ F(t'+i\beta/4)\)}{\int_{t}^\infty dt' e^{-\frac{2\pi  (2J+1)}{\beta}t'}\(F_d-\mbox{Re}\ F(t'+i\beta/4)\)} < 1
\end{align}
since $F_d-\mbox{Re}\ F(t'+i\beta/4)> 0$. Moreover, $\mu_{J+1}\(t\)=\mu_{J}\(t\)$, for any $t>t_0$ and integer $J\ge 0$ necessarily requires that  $\mu_J\(t\)=0$ for all integer $J$ implying the system is non-chaotic. 

So, for any chaotic system we conclude that moments $\mu_{J}\(t\)$, as a function of integer $J\ge 0$, must be strictly monotonically decreasing, positive, and bounded above the factorization time $t\ge t_0$.\footnote{The factorization assumption (\ref{eq:factorization}), as mentioned in \cite{Maldacena:2015waa}, can break down for $t\gg t_*$ due to Poincare recurrences. So, strictly speaking, the bounds of this paper, just like the MSS bound, are valid up to some cut-off scale which is a lot larger than the scrambling time, but smaller than the Poincare recurrence time of the system.  } Later, we will show that these conditions contain the MSS chaos bound. 

\subsubsection{Log-Convexity Condition}
We can  derive a stronger but non-linear constraint on $\mu_J(t)$. For all integer $J\ge 0$, the Cauchy-Schwarz inequality of real integrable functions imposes
\be
0< \frac{\(\int_{t}^\infty dt' e^{-\frac{2\pi  (2J+3)}{\beta}t'}\(F_d-\mbox{Re}\ F(t'+i\beta/4)\)\)^2}{\int_{t}^\infty dt' e^{-\frac{2\pi  (2J+1)}{\beta}t'}\(F_d-\mbox{Re}\ F(t'+i\beta/4)\)\int_{t}^\infty dt'' e^{-\frac{2\pi  (2J+5)}{\beta}t''}\(F_d-\mbox{Re}\ F(t'+i\beta/4)\)}\le 1\ ,
\ee
where $t\ge t_0$. Now using the relation (\ref{eq:dispersion}) we obtain $\mu_{J+1}(t)^2\le \mu_J\(t\)\mu_{J+2}\(t\)$ for any $t\ge t_0$ and integer $J\ge 0$. This ``local" (in $J$) condition can be applied iteratively to obtain a global condition
\be\label{eq:superconvexity}
\frac{1}{J_2-J_1}\ln \(\frac{\mu_{J_1}(t)}{\mu_{J_2}(t)}\)\ge \frac{1}{J_3-J_1}\ln \(\frac{\mu_{J_1}(t)}{\mu_{J_3}(t)}\)\ , \qquad 0\le J_1<J_2<J_3
\ee
for all integer $J_1, J_2$, and $J_3$ implying that moments $\mu_{J}\(t\)$, for non-negative integer $J$, form a logarithmically convex (also known as super-convex) function of $J$ above the factorization time $t\ge t_0$.\footnote{Of course, this implies that $\mu_{J}\(t\)$, as a function of integer $J$, is also convex. However, log-convexity condition is a stronger restriction on $\mu_{J}\(t\)$. Note that the above log-convexity condition is closely related to the Hadamard three-lines theorem.} 

It is known that CFT Regge correlators obey similar positivity, monotonicity, and log-convexity conditions, as shown in \cite{Kundu:2020gkz,Kundu:2021qpi}. At this point this should not be a surprise.  In fact, in section \ref{sec:CFT} we will relate the above bounds on $\mu_J(t)$ with the CFT bounds of \cite{Kundu:2020gkz,Kundu:2021qpi} by interpreting certain CFT four-point correlators as OTOCs in a thermal state  on Rindler space.

\subsubsection{Bounds on the Rate of Change}
There is also a bound on the rate of change of moments $\mu_J(t)$. From (\ref{eq:dispersion}) we find that 
\be\label{eq:rate}
e^{\frac{4\pi J t}{\beta}}\ \frac{d}{dt} \(e^{-\frac{4\pi J t}{\beta}}\mu_J(t)\)=-2 e^{-\frac{2\pi  }{\beta}t}\(F_d-\mbox{Re}\ F(t+i\beta/4)\)
\ee
for any real $t\ge t_0$ and integer $J$ (positive or negative).\footnote{When $J$ is a negative integer, equation (\ref{eq:rate}) can be derived by taking the $\gamma$-contour to be between $t$ and $t+\delta t$.} First of all, the right-hand side is independent of $J$. Secondly, the property (\ref{eq:positive}) imposes
\be\label{eq:ratebound}
\frac{d \mu_J(t)}{dt}\le \frac{4\pi J}{\beta} \mu_J(t)\qquad \text{for} \qquad t\ge t_0
\ee
and all integer $J$. The above condition is very similar to the MSS condition (\ref{bound:MSS}), however, they are not exactly equivalent. Note that the primary moment $\mu_0(t)$ is a monotonically decreasing function of time above the factorization time $t\ge t_0$.

The moments $\mu_J(t)$, for non-integer $J$, are less interesting since they are weakly constrained. However, they are not free of constraints, as we discuss in appendix \ref{app:noninteger}.

There are some chaotic systems in which the OTOC is known beyond the leading Lyapunov term. For example, subleading corrections of the OTOC (\ref{eq:otoc})  are known in 2d CFT \cite{Roberts:2014ifa}, as well as in the Schwarzian theory \cite{Maldacena:2016upp,Lam:2018pvp}. In both these cases, one can analytically check that OTOCs satisfy all constraints (\ref{intro:positive})-(\ref{intro:log}) for $|t-t_*|\gg \frac{\beta}{2\pi}$. Near the scrambling time scale, one can check these constraints numerically.

\subsection{OTOC as a Sum Over Moments}
Consider a contour integral 
\be
\oint_{\gamma'} dt' \ e^{\frac{2\pi }{\beta}(t'-t)} \frac{F(t')-F_d}{1-e^{\frac{4\pi }{\beta}(t'-t)}}=\frac{i \beta}{2}  (F(t)-F_d) \ ,
\ee
where $|\mbox{Im}\ t| <\beta/4$ and the contour $\gamma'$ is chosen to be a small rectangle around the point $t$ with four sides at $\mbox{Re}\ t'=\mbox{Re}\ t\pm \epsilon$ and $\mbox{Im}\ t=\pm \beta/4$. In the limit $\epsilon\rightarrow 0$, we obtain 
\begin{align}
\frac{i \beta}{2}  (F(t)-F_d)=&\int_{{\rm Re} (t)-\epsilon-i \frac{\beta}{4}}^{{\rm Re}(t)-\epsilon+i \frac{\beta}{4}} dt'  e^{\frac{2\pi }{\beta}(t'-t)} \frac{F(t')-F_d}{1-e^{\frac{4\pi }{\beta}(t'-t)}}\nonumber\\
&
-\int_{{\rm Re} (t)+\epsilon-i \frac{\beta}{4}}^{{\rm Re}(t)+\epsilon+i \frac{\beta}{4}} dt'  e^{\frac{2\pi }{\beta}(t'-t)} \frac{F(t')-F_d}{1-e^{\frac{4\pi }{\beta}(t'-t)}}+ I\ ,
\end{align}
where, $I$ is the integral on the edges $\mbox{Im}\ t=\pm \beta/4$, which is bounded $|I| \le |F_d-F_+(t)|\epsilon \rightarrow 0$.
Therefore, we can write 
\begin{align}
\frac{ \beta}{2}  (F_d-F(t))=&-\int_{{\rm Re} (t)-\epsilon-i \frac{\beta}{4}}^{{\rm Re}(t)-\epsilon+i \frac{\beta}{4}} dt'  \sum_{n=0}^\infty (-1)^n e^{\frac{2\pi }{\beta}(t'-t)(2n+1)-i(2n+1)\frac{\pi}{2}}\(F(t')-F_d\)\nonumber\\
& +\int_{{\rm Re}(t)+\epsilon-i \frac{\beta}{4}}^{{\rm Re}(t)+\epsilon+i \frac{\beta}{4}} dt'  \sum_{n=0}^\infty (-1)^n e^{\frac{2\pi }{\beta}(t-t')(2n+1)+i(2n+1)\frac{\pi}{2}}\(F(t')-F_d\)
\end{align}
provided $\epsilon\rightarrow 0_+$.  The above relation can be rewritten as  
\be\label{eq:master}
F(t)=F_d-\frac{2}{\beta}e^{\frac{2\pi}{\beta}t}\sum_{J=0,\pm 1,\pm 2, \cdots} (-1)^{J}\mu_J(\mbox{Re}\ t) e^{\frac{4\pi J }{\beta}i{\rm Im}(t)}\ , \qquad |\mbox{Im}\ t |< \frac{\beta}{4}\ .
\ee
So, if we know all the integer moments at any $t$, we also know the OTOC exactly at that point of time. This is a non-trivial formula which we have checked explicitly for simple cases. 

Notice that for $|\mbox{Im}\ t |= \frac{\beta}{4}$, the expansion (\ref{eq:master}) leads to $\mbox{Re}(F(t)-F_d)=0$. However, for any chaotic system $\mbox{Re}(F(t)-F_d)$ cannot be identically zero on the boundary $|\mbox{Im}\ t |= \frac{\beta}{4}$ of the half-strip (\ref{halfstrip}). This is not actually a contradiction since the expansion (\ref{eq:master}) breaks down for $|\mbox{Im}\ t |= \frac{\beta}{4}$. A similar expansion can be written down for $|\mbox{Im}\ t |= \frac{\beta}{4}$ which we will discuss in appendix \ref{app:boundary}.

\subsection{Late-Time Behavior}
The expansion (\ref{eq:master}), at late times ($t\gg t_0$), is actually equivalent to the dispersion relation (\ref{OTOC:sumrule}). This follows from the fact that the late-time behavior of $F(t)$ is completely fixed by the primary moment $\mu_0(t)$.  In order to show this first note that for any integer $J\ge 0$, equation (\ref{eq:rate}) implies 
\be\label{eq:consistency}
\mu_J( t)=- e^{\frac{4\pi J}{\beta}t}\int_{t}^\infty dt' e^{-\frac{4\pi J}{\beta}t'} \mu_0'(t')\ ,
\ee
where we have used the fact that $\mu_J(t\rightarrow \infty)=0$ when $J$ is a non-negative integer. Moreover, for negative moments we find that 
\be
\mu_{-J}( t)=\mu_{-J}(t_0)e^{-\frac{4\pi J}{\beta}(t-t_0)}+e^{-\frac{4\pi J}{\beta}t} \int^{t}_{t_0} dt' e^{\frac{4\pi J}{\beta}t'} \mu_0'(t')
\ee
where $J$ is a positive integer. Notice that  the first term on the right-hand side decays at late times $t\gg t_0$. We now use the expansion (\ref{eq:master}) to obtain 
\begin{align}
F_d-F(t)=\frac{2}{\beta}e^{\frac{2\pi}{\beta}t}\sum_{J= 1}^\infty (-1)^{J} \(  \mu_{-J}(t_0) e^{-\frac{4\pi J}{\beta}(t-t_0)}+e^{-\frac{4\pi J}{\beta}t} \int^{{\rm Re} t}_{t_0} dt' e^{\frac{4\pi J}{\beta}t'} \mu_0'(t') \right)\nonumber\\
-\frac{2}{\beta}e^{\frac{2\pi}{\beta}t}\sum_{J= 0}^\infty (-1)^{J} e^{\frac{4\pi J}{\beta}t} \int_{{\rm Re} t}^\infty dt' e^{-\frac{4\pi J}{\beta}t'}  \mu_0'(t') 
\end{align}
for $|\mbox{Im}\ t |< \frac{\beta}{4}$. We sum over the last two terms, yielding\footnote{This equation will play an important role for extremally chaotic OTOCs in \cite{Kundu:2021mex}.}
 \begin{align}\label{eq:late}
F_d-F(t)= \frac{2}{\beta}e^{\frac{2\pi}{\beta}t}\sum_{J= 1}^\infty (-1)^{J} \mu_{-J}(t_0) e^{-\frac{4\pi J}{\beta}(t-t_0)}- \frac{2}{\beta} e^{\frac{2\pi}{\beta}t}\int_{t_0}^\infty dt' \frac{\mu_0'(t')}{1+e^{\frac{4\pi }{\beta}(t-t')}}\ .
\end{align}
The first term decays for ${\rm Re}\ t\gg t_0$. Hence, the late-time behavior is controlled completely by the primary moment $\mu_0(t)$.  Furthermore, note that the above equation is exactly the same as (\ref{OTOC:sumrule}), once we use (\ref{eq:rate}).

In terms of the primary moment, maximal chaos is defined as follows. There is some time interval $t_0\ll t_i\le t\le t_f$ in which $\mu_0(t)\approx$ constant, where $t_f\gg t_i$. In this case, as one can see from equation (\ref{eq:late}) that
\be
F(t)\approx F_d-\frac{2\mu_0(t_f)}{\beta}e^{\frac{2\pi}{\beta}t}
\ee
for $t_i\le t\ll t_f$. However, note that analyticity does not allow $\mu_0(t)$ to be exactly a constant over any finite time interval. Hence, a Lyapunov growth that saturates the MSS bound must always come with correction terms. We will discuss these correction terms in the next section.

\section{Implications of the Bounds}\label{sec:implications}
\subsection{Separation of Scales and MSS Bound}\label{sec:mss}
Bounds of the previous section are rather general, since they follow from the assumptions of section \ref{sec:assumptions}.  So far, we have not assumed anything about the form of $F(t)$. When the scrambling time is much larger than the other time scales $t_* \gg t_0, t_d$, the OTOC $F(t)$ is expected to exhibit a period of exponential growth \cite{Liao:2018uxa,Lewis-Swan:2019pln}  characterized by 
\be\label{eq:growth}
F_d - F(t)= \frac{c_1}{\N}e^{\lambda_L t}+\cdots \ , \qquad t_* \gg t\gg  t_0
\ee
where, $\lambda_L>0$ is the Lyapunov exponent, $c_1$ is an order one constant, and $\N$ is the effective density of degrees of freedom. Moreover, for such a system, the scrambling time \cite{Sekino:2008he} is defined as
\be
t_*= \frac{1}{\lambda_L}\ln \N
\ee
such that $F_d - F(t)=c_1 \exp(\lambda_L(t-t_*))$.   It is a straightforward exercise to compute the moments by using the definition (\ref{def:moments}):
\be\label{moments:lyapunov}
\mu_J(t)=\frac{1}{\N} \frac{2\beta c_1}{2 \pi  (2J+1)-\beta  \lambda_L}  e^{t \left(-\frac{2 \pi  }{\beta }+\lambda_L \right)}\cos \left(\frac{\beta  \lambda_L }{4}\right)
\ee
for integer $J$.\footnote{When $\frac{\beta \lambda_L}{2\pi}$ is an odd integer, one should calculate moments  by taking limits in (\ref{moments:lyapunov}).} Notice that $\mu_J(t)$, for $\lambda_L >\frac{2\pi}{\beta}$,  changes sign as we move from $J<(\frac{\beta \lambda_L}{2\pi}-1)/2$ to $J>(\frac{\beta \lambda_L}{2\pi}-1)/2$, provided $\lambda_L \neq \frac{2\pi}{\beta}(2n+1)$ with integer $n$. This is inconsistent with the positivity condition (\ref{eq:dispersion}). On the other hand, for $\lambda_L = \frac{2\pi}{\beta}(2n+1)$ with integer $n$, all $\mu_{J\neq n}(t)=0$, up to subleading terms in (\ref{eq:growth}). This violates the monotonicity condition (\ref{eq:monotonicity}) when $n\ge 1$. Hence, positivity and monotonicity  of $\mu_J(t)$ for all integer $J\ge 0$ imply the MSS bound on the  Lyapunov exponent
\be
\lambda_L \le \frac{2\pi}{\beta}\ .
\ee
Moreover, the condition (\ref{eq:dispersion}) also requires that $c_1>0$. Note that (\ref{moments:lyapunov}) now automatically satisfies the log-convexity condition (\ref{eq:superconvexity}). Whereas, the boundedness condition (\ref{eq:bounded}) tells us when the approximation (\ref{eq:growth}) breaks down. 

\subsection{Bounds on Subleading Corrections}
We now explore bounds on correction terms characterized by 
\be\label{otoc:two}
F_d - F(t)= \frac{c_1}{\N}e^{\lambda_1 t}+ \frac{c_2}{\N}e^{\lambda_2 t} + \cdots \ , \qquad t_f \ge \mbox{Re}\ t\ge  t_i
\ee
where, $\lambda$-exponents are positive and $\lambda_2>\lambda_1$. The above parametrization is valid in some regime $t_f \ge \mbox{Re}\ t\ge  t_i$ where $t_0< t_i\ll t_f <t_*$ . To begin with, we only  assume that $c_1\sim \O(1)$, however, we will not make the same assumption about the $c_2$ coefficient. In other words, even though the second term of (\ref{otoc:two}) grows faster, we are not making any assumption whether and in what range it dominates.\footnote{For example, $|c_2|/|c_1|$ can be suppressed by some positive powers of $1/\N$.}

We again calculate moments by using the definition (\ref{def:moments}):
\be\label{eq:twocoeff}
\mu_J(t)=\frac{2\beta}{\N} e^{-\frac{2 \pi  }{\beta }t} \(\frac{ c_1 e^{\lambda_1 t }\cos \left(\frac{\beta  \lambda_1 }{4}\right)}{2 \pi  (2J+1)-\beta  \lambda_1}+\frac{ c_2 e^{\lambda_2 t }\cos \left(\frac{\beta  \lambda_2 }{4}\right)}{2 \pi  (2J+1)-\beta  \lambda_2}  \)+\cdots 
\ee
for $t_f \ge t\ge  t_i$ and  integer $J$. From the above expression, it is easy to argue that $\lambda_1\le \frac{2\pi}{\beta}$. First, note that both terms in (\ref{eq:twocoeff}) cannot dominate within the entire regime $t_f \ge t\ge  t_i$, since $\lambda_2>\lambda_1$. So, there must be some duration within $t_f \ge t\ge  t_i$ in which either the first term or the second term dominates. We can now repeat the argument of section \ref{sec:mss} for the dominant contribution. Both scenarios necessarily require 
\be
\lambda_1\le \frac{2\pi}{\beta}\ .
\ee
However, in general the same bound does not apply to $\lambda_2$ because it is possible that the second term never dominates in the duration $t_f \ge t\ge  t_i$.

\subsubsection{$\lambda_2> \frac{2\pi}{\beta}$}
Now we fix $\lambda_1< \frac{2\pi}{\beta}$ and argue that $\lambda_2> \frac{2\pi}{\beta}$ is allowed as long as $|c_2|/|c_1|$ is fined-tuned.\footnote{Maximally chaotic systems saturate the MSS bound $\lambda_1= \frac{2\pi}{\beta}$. This case will be discussed separately. } Of course, the preceding discussion implies that the  second term in (\ref{eq:twocoeff}), for any integer $J\ge 0$, can never dominate in the regime $t_f \ge t\ge  t_i$. This immediately implies 
\be
c_1>0 \ .
\ee
Moreover, the MSS bound (\ref{intro:MSS}) at $t=t_f$ imposes
\be\label{two:MSS}
-e^{-(\lambda_2-\lambda_1)t_f}\le \frac{c_2}{c_1}\le \(\frac{2\pi-\lambda_1 \beta}{\lambda_2 \beta-2\pi}\) e^{-(\lambda_2-\lambda_1)t_f}\ ,
\ee
where the lower bound is simply the statement that $F(t_f)\le F_d$. Hence, $c_2/c_1$ must be exponentially suppressed $e^{-(\lambda_2-\lambda_1)t_f}$.

\begin{figure}
\centering
\includegraphics[scale=0.58]{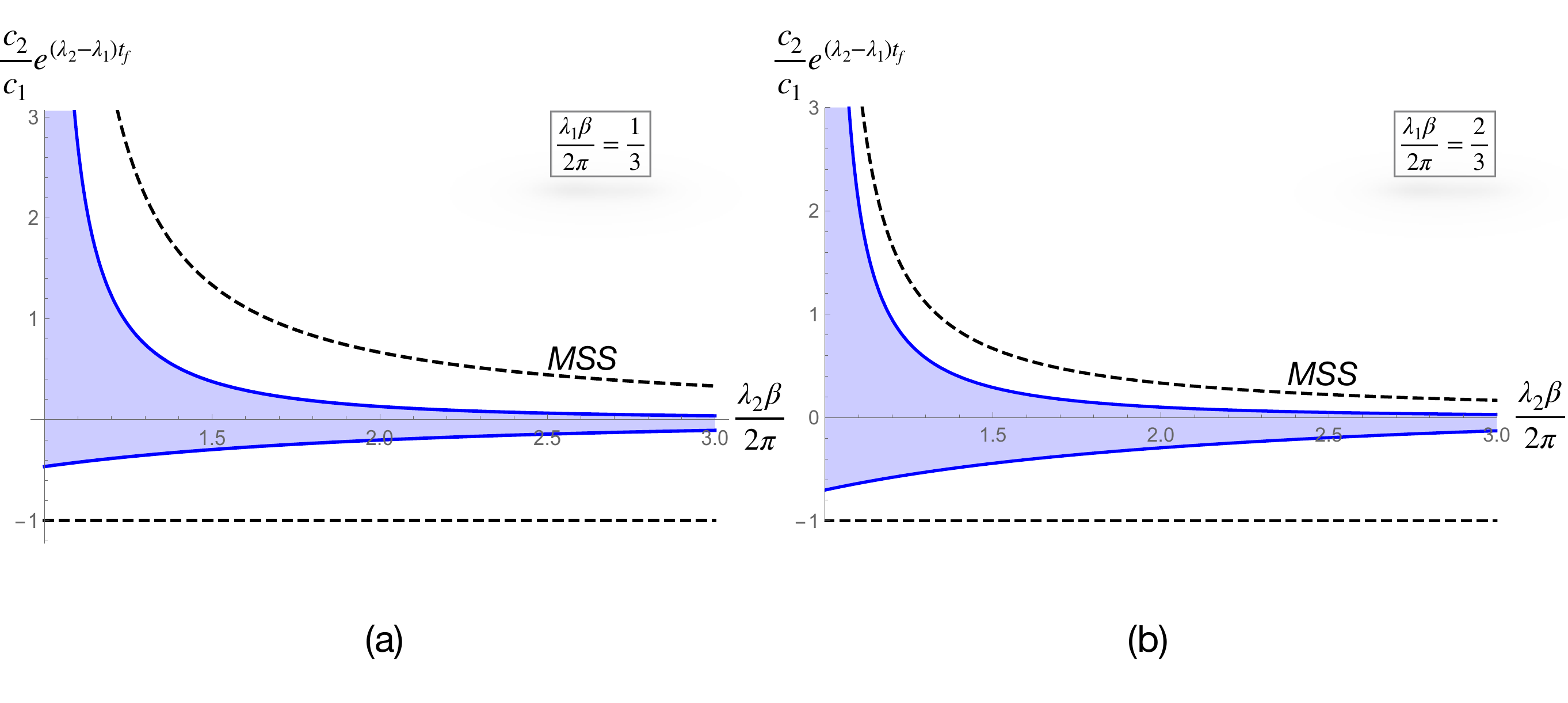}
\caption{ \label{fig:bound} \small Bounds on subleading corrections for $\lambda_1< \frac{2\pi}{\beta}$ and $\frac{2\pi}{\beta}<\lambda_2< \frac{6\pi}{\beta}$. The chaos bounds necessarily require that $c_1>0$ and $c_2$ is parametrically suppressed. Furthermore, the chaos bounds of section \ref{sec:BoundsOnMoments} impose two-sided bounds on $c_2/c_1$ as shown here (blue lines) for (a) $\frac{\lambda_1\beta}{2\pi}=\frac{1}{3}$ and (b) $\frac{\lambda_1\beta}{2\pi}=\frac{2}{3}$. In particular, the unshaded regions are ruled out by these bounds. Black dashed lines represent constraints from the MSS bound (\ref{two:MSS}) which are strictly weaker. }
\end{figure}

The chaos bounds of the preceding section impose stronger restriction on $c_2 e^{(\lambda_2-\lambda_1)t_f}/c_1$ than that from the MSS bound (\ref{two:MSS}). The bounds of this paper necessarily require
\be\label{eq:twobounds}
\mu_J(t_f)>0\ ,\qquad \mu_J(t_f)-\mu_{J+1}(t_f)>0\ , \qquad \mu_{J}(t_f)\mu_{J+2}(t_f)-\mu_{J+1}(t_f)^2\ge 0
\ee 
for all  integer $J\ge 0$, where $\mu_J(t)$ is given by (\ref{eq:twocoeff}). In order to show that these bounds are stronger than (\ref{two:MSS}), we consider the example $\frac{6\pi}{\beta}>\lambda_2> \frac{2\pi}{\beta}$. In this case, it is sufficient to study $J=0$ and $1$ of (\ref{eq:twobounds}). In particular, a careful consideration reveals that the optimal bounds on $c_2/c_1$ are obtained from
\begin{align}
\text{Lower bound:}\qquad \mu_{0}(t_f)\mu_{2}(t_f)-\mu_{1}(t_f)^2\ge 0\ , \label{sol1}\\
\text{Upper bound:}\qquad \mu_{1}(t_f)\mu_{3}(t_f)-\mu_{2}(t_f)^2\ge 0\ , \label{sol2}
\end{align}
which can be solved analytically. The actual solutions are not very illuminating and hence we will not transcribe them here. Rather, we plot the bounds obtained from (\ref{sol1}) and (\ref{sol2}) in figure  \ref{fig:bound}  and compare them with results from the MSS bound (\ref{two:MSS}). Of course, a similar analysis can also be performed even when $\lambda_2\ge \frac{6\pi}{\beta}$ with qualitatively similar results.

\subsubsection{$\lambda_1<\lambda_2\le \frac{2\pi}{\beta}$}
This scenario is less constrained. The region $c_1, c_2>0$ is obviously allowed. On the other hand, the region $c_1, c_2<0$ is ruled out. However, now one of the $c$-coefficients can be a little negative. 

In this case, bounds on $c_1$ and $c_2$ depend on which of the contribution in (\ref{otoc:two}) dominates. As before, if the first term of (\ref{eq:twocoeff}) dominates for all integer $J\ge 0$  in the entire duration $t_f \ge t\ge  t_i$, then there is no bound on the the second term. In other words
\be\label{eq:region1}
\text{if} \qquad \frac{|c_2|}{|c_1|}\le  e^{-(\lambda_2-\lambda_1)t_f} \frac{\cos \left(\frac{\beta  \lambda_1 }{4}\right)}{\cos \left(\frac{\beta  \lambda_2 }{4}\right)}\(\frac{2\pi-\beta \lambda_2}{2\pi-\beta \lambda_1}\)\ \Rightarrow\qquad c_1>0
\ee
but $c_2$ is allowed to be negative. 

Similarly, the second term dominates within the entire regime $t_f \ge t\ge  t_i$
\be\label{eq:region2}
\text{if} \qquad \frac{|c_2|}{|c_1|}\ge  e^{-(\lambda_2-\lambda_1)t_i} \frac{\cos \left(\frac{\beta  \lambda_1 }{4}\right)}{\cos \left(\frac{\beta  \lambda_2 }{4}\right)}\ \Rightarrow \qquad c_2>0
\ee
but now there is no sign constraint on $c_1$. In the above expression, we have assumed that $\lambda_2<\frac{2\pi}{\beta}$.\footnote{In the special case, $\lambda_2=\frac{2\pi}{\beta}$, equation (\ref{eq:region2}) becomes 
\be
\text{if} \qquad \frac{|c_2|}{|c_1|}\ge 4 e^{-(\lambda_2-\lambda_1)t_i} \frac{\cos \left(\frac{\beta  \lambda_1 }{4}\right)}{2\pi-\beta \lambda_1}\ \Rightarrow \qquad c_2>0\ .
\ee
The case of maximal chaos will be discussed in section \ref{sec:maximal} with more details. }

Of course, if $|c_2/c_1|$ is in between regions (\ref{eq:region1}) and (\ref{eq:region2}), then both of them must be positive
\be
c_1>0 \ , \qquad c_2>0\ .
\ee

\subsection{Maximal Chaos and Subleading Terms}\label{sec:maximal}
It is well-known that large $N$ theories that are holographically dual to Einstein gravity saturate the MSS bound. For these theories, the OTOC can be well-approximated by (\ref{eq:growth}) with $\lambda_L=\frac{2\pi}{\beta}$ and the associated moments can be computed from (\ref{moments:lyapunov}). In particular, for integer $J$ we obtain
\begin{align}\label{qg:moments}
\mu_0(t)=\frac{c_1 \beta}{2\N}\ , \qquad \mu_{J\ge 1}(t)=0
\end{align}
implying $c_1>0$. However, we immediately see that (\ref{qg:moments}) is inconsistent with the sum-rule (\ref{eq:dispersion}) which requires all integer $J\ge 0$ moments $\mu_{J}(t)$ to be strictly positive since $\mu_0(t)>0$. This simply means that there has to be subleading corrections to the leading Lyapunov growth. Let us parametrize the subleading corrections as follows:
\be\label{eq:para}
F_d - F(t)= \frac{1}{\N}\(c_1e^{\frac{2\pi}{\beta} t}+ c_2 \varepsilon e^{\lambda_2 t}+\cdots\) \ ,
\ee
where, $c_1,c_2\sim \O(1)$ and  $\varepsilon$ is a small expansion parameter which is positive. For example, $\varepsilon$ can simply be some powers of $\frac{1}{\N}$. Of course, in general, it can also be a completely independent parameter. For either case, moments can be determined in the same way
\be
\mu_J(t)=\frac{\beta}{\N}\(\frac{c_1 }{2} \delta_{J,0}+ \frac{2 c_2 \varepsilon}{2 \pi  (2J+1)-\beta  \lambda_2}  e^{t \left(-\frac{2 \pi  }{\beta }+\lambda_2 \right)}\cos \left(\frac{\beta  \lambda_2}{4}\right)\)
\ee
where $J\ge 0$ is an integer. The positivity and the monotonicity conditions of the previous section impose
\be\label{bound:2}
\lambda_2 \le \frac{6\pi}{\beta}\ , \qquad c_2 \cos \left(\frac{\beta  \lambda_2}{4}\right) \ge 0 \ .
\ee
So, $\lambda_2$ is allowed to be larger than $\frac{2\pi}{\beta}$. 

The OTOC must not violate the log-convexity condition even if $\lambda_2 > \frac{2\pi}{\beta}$. In particular, the condition (\ref{intro:log}), for $J=0$, imposes 
\be\label{max:c2bound}
c_2 \varepsilon e^{t \left(\lambda_2-\frac{2 \pi  }{\beta } \right)}\ge \frac{c_1 }{(8\pi)^2}\frac{(\beta  \lambda _2-2 \pi) (6 \pi-\beta  \lambda_2  )^2}{\cos \left(\frac{\beta  \lambda_2}{4}\right)}
\ee
when $\lambda_2 > \frac{2\pi}{\beta}$. The above inequality should be interpreted as an upper bound on $t$ above which the parametrization (\ref{eq:para}) breaks down for $\frac{6\pi}{\beta}> \lambda_2 > \frac{2\pi}{\beta}$. Interestingly, the right-hand side vanishes when $\lambda_2$ also saturates the subleading chaos bound (\ref{bound:intro}). In that case, all the chaos bounds are automatically satisfied, provided $c_2<0$. 

Let us now further explore the case in which the OTOC (\ref{eq:para}) saturates the subleading chaos bounds (\ref{bound:intro}) as well. This fixes $\lambda_2 = \frac{6\pi}{\beta}$. It is tempting to ask whether the same pattern persists even for higher order correction terms. For example, we can add another exponential (sub-subleading) correction term 
\be
 \frac{\varepsilon^2}{\N} c_3 e^{\lambda_3 t}
\ee
to  (\ref{eq:para}). We repeat the preceding argument which now imposes
\be
\lambda_3 \le \frac{10\pi}{\beta}\ , \qquad c_3 \cos \left(\frac{\beta  \lambda_3}{4}\right) \ge 0\ .
\ee
It is easy to see that this argument generalizes for all higher-order correction terms. In particular, if first $n$ Lyapunov exponents saturate the condition $\lambda_n =(2n-1)\frac{2\pi}{\beta}$, then there is a bound on the $(n+1)$-th Lyapunov exponent
\be
\lambda_{n+1} \le(2n+1)\frac{2\pi}{\beta}\ .
\ee
Recently, a hydrodynamic effective field theory of chaos has been proposed to describe maximally chaotic systems \cite{Blake:2017ris, Blake:2021wqj}.\footnote{This is related to pole-skipping in the energy-density two-point function \cite{Blake:2017ris,Grozdanov:2017ajz,Haehl:2018izb,Blake:2018leo,Grozdanov:2018kkt,Haehl:2019eae,Ahn:2019rnq,Ahn:2020bks,Ramirez:2020qer,Choi:2020tdj}.} It would be interesting to understand the bounds of this paper from this effective field theory perspective. 

\subsubsection{Example: Schwarzian Theory}
Many key features of 2D black holes, such as strong chaos, are effectively described by the Schwarzian theory, which is maximally chaotic. Moreover, in the Schwarzian theory, it is possible to compute the OTOC (\ref{eq:otoc}) as an expansion in $1/\N$.  In particular, the OTOC is given by  the confluent hypergeometric $U$-function \cite{Maldacena:2016upp,Lam:2018pvp} which can be expanded, yielding 
\be\label{otoc:sch}
F_d-F(t)=F_d \(\frac{  \Delta ^2 }{4 \pi }e^{\frac{2 \pi  (t-t_*)}{\beta }}-\frac{ (2 \Delta +1)^2 \Delta ^2 }{128 \pi ^2}e^{\frac{4 \pi  (t-t_*)}{\beta }}+\cdots\)\qquad  t<t_f\ ,
\ee
where $\Delta$ is the scaling dimension of both operators $V$ and $W$.  The higher-order terms can only be ignored for $t<t_f$, where $t_f=t_*+ \frac{\beta}{2\pi} \ln \frac{12 \pi }{(\Delta +1)^2}$. Specifically, $t_f$ is computed by comparing the subleading term of  (\ref{otoc:sch}) with the next order term. 

The above OTOC has the exact form (\ref{eq:para}). First, we notice that the bound (\ref{bound:2}) is satisfied by the OTOC (\ref{otoc:sch}). We now focus on the bound (\ref{max:c2bound}) which simplifies to 
\be
- \frac{c_2 \varepsilon}{c_1}e^{\frac{2\pi}{\beta} t} \le \pi\ .
\ee
The OTOC (\ref{otoc:sch}) is consistent with this bound even at $t=t_f$ for all $\Delta> 0$.

\subsection{A Special Case}\label{sec:special}
At this stage, what could be more natural than to consider the special case\footnote{We are writing $\varepsilon=e^{-\frac{4\pi}{\beta}t_f}$ in (\ref{eq:specialcase}), where $t_f$ is not required to be parametrically smaller than the scrambling time $t_*$.  }
\be\label{eq:specialcase}
F_d - F(t)= \frac{1}{\N}e^{\frac{2\pi}{\beta} t}\sum_{n=0}^{\infty}c_{n+1} e^{\frac{4n\pi}{\beta} (t-t_f)}\ , \qquad  t_f\ge \mbox{Re}\ t\gg t_0
\ee 
which we assume to converge for some $t_f\gg t_0$.  The above OTOC saturates (\ref{eq:cc3}) exactly for $N\rightarrow \infty$.\footnote{Note that the constraint (\ref{eq:cc3}) can be exactly saturated only for $N=\infty$.}  Furthermore, bounds  from section \ref{sec:BoundsOnMoments}  impose additional precise constraints, beyond the consistency conditions (\ref{eq:cc3}). To utilize the full set of bounds from section \ref{sec:BoundsOnMoments}, we first compute moments $\mu_J(t)$ from the OTOC (\ref{eq:specialcase})
\be
\mu_{J}(t)=(-1)^J\frac{\beta }{2\N}c_{J+1}e^{\frac{4J\pi}{\beta}(t-t_f)}\ , \qquad J=0,1,2,\cdots
\ee
for $t_0\ll t\le  t_f$. So, the positivity condition (\ref{eq:dispersion}) imposes 
\be\label{c:bound1}
(-1)^n c_{n+1}>0\ .
\ee
Besides, the monotonicity condition (\ref{eq:monotonicity}), for $t_0\ll t\le t_f$, necessarily requires that  
\be
|c_{n+1}|<|c_n|\ , \qquad n=1,2,3,\cdots\ . 
\ee
This condition implies that the term  that grows as $e^{\frac{2\pi}{\beta} t}$ must always dominate over the rest of the higher-order terms within the regime of validity. This is precisely the statement that the OTOC cannot grow faster than $e^{\frac{2\pi}{\beta} t}$.

Finally the log-convexity condition (\ref{eq:superconvexity}) imposes 
\be\label{c:bound3}
c_{n+1}^2\le c_n c_{n+2}\ , \qquad n=1,2,3,\cdots\ . 
\ee
On the other hand, the bound (\ref{eq:ratebound}) on the rate of change is automatically saturated by the OTOC (\ref{eq:specialcase}). 

So, the $c$-coefficients of the OTOC (\ref{eq:specialcase}) must satisfy the conditions (\ref{c:bound1})-(\ref{c:bound3}). Let us emphasize that among these conditions, only the condition (\ref{c:bound3}) can be saturated by a quantum system. This fact will play a central role in the discussion of extremal chaos in \cite{Kundu:2021mex}. This special case also has applications in CFT, as we will discuss in the next section.
 
 \section{Conformal Regge Correlators: A Consistency Check}\label{sec:CFT}
 There is an apparent similarity between conditions (\ref{intro:positive})-(\ref{intro:log}) on moments of OTOCs  and analogous positivity, monotonicity, and log-convexity conditions of  CFT Regge correlators, as obtained in \cite{Kundu:2020gkz,Kundu:2021qpi} from basic properties of Lorentzian correlators. We now relate these two sets of bounds by following \cite{Maldacena:2015waa}. 
 
\begin{figure}
\begin{center}
\begin{tikzpicture}[baseline=-3pt, scale=1.80]
\begin{scope}[very thick,shift={(4,0)}]
\coordinate (v1) at (-1.5,-1.5) {};
\coordinate(v2) at (1.5,1.5) {};
\coordinate (v3) at (1.5,-1.5) {};
\coordinate(v4) at (-1.5,1.5) {};
\draw[thin,-latex]  (v1) -- (v2)node[left]{$x^+$};
\draw[thin,-latex]  (v3) -- (v4)node[right]{$\ x^-$};
\draw(-1.6,0)node[left]{ $\ V(1,-1)$};
\draw(1.6,0)node[right]{ $ V(-1,1)$};
\filldraw[black]  (-1.6,0) circle (1pt);
\filldraw[black]  (1.6,0) circle (1pt);
\coordinate(v5) at (0,0) {};
\def \fac {.6};
\filldraw[black]  (-1.2,1) circle (1 pt);
\filldraw[black]  (1.2,-1) circle (1pt);
\draw(-1.2,1)node[left]{ $ W(\rho,-\bar{\rho})$};
\draw(1.2,-1)node[right]{ $ W(-\rho,\bar{\rho})$};
\end{scope}
\end{tikzpicture}
\end{center}
\caption{\label{config} \small A Lorentzian four-point function where all points are restricted to a $2$d subspace $\{x^0,x^1\}$. Null coordinates are defined as $x^{\pm}=x^0\pm x^1$. }
\end{figure}
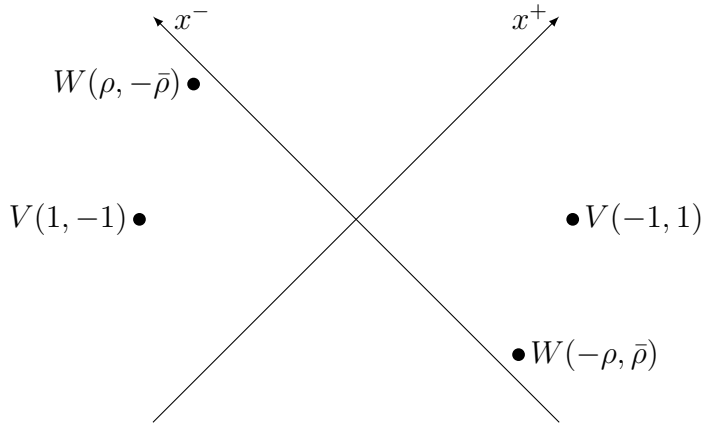

Consider a CFT$_d$ Lorentzian correlator 
\be\label{eq:correlator}
G=\frac{\langle V(x_4) W(x_1) W(x_2) V(x_3)  \rangle}{\langle  W(x_1) W(x_2)   \rangle\langle V(x_4) V(x_3)  \rangle}
\ee
in the kinematics
\begin{align}\label{points}
x_1=-x_2=(x^-=\rho,x^+=-\bar{\rho},\vec{0})\ ,  \qquad x_3=-x_4=(x^-=-1,x^+=1,\vec{0})\ ,
\end{align}
with $0<\rho<1$ and $\br>1$, as shown in figure \ref{config}, where operators inside the correlator are ordered as written. We are using null coordinates $\{x^{\pm},\vec{x}_{d-2}\}\equiv\{x^0\pm x^1,\vec{x}_{d-2}\}$, where $x^0$ is time. In order to compare with \cite{Kundu:2020gkz,Kundu:2021qpi}, we  parametrize 
\be\label{sigma}
\r=\frac{1}{\sigma}\ , \qquad \br=\sigma \eta  
\ee
with $\eta>0$ and hence $G\equiv G(\eta,\sigma)$.

The Minkowski vacuum can be equivalently described as the thermofield double, entangling the right Rindler wedge with the left Rindler wedge. In this description the Lorentzian correlator  $G$ can be interpreted as a thermal correlator where $T=1/\beta$ is the Rindler temperature. More precisely, we choose Rindler coordinates 
\begin{align}
&\{x^{-},x^+,\vec{x}_{d-2}\}_L=\{re^{\frac{2\pi}{\beta} t_R},-r e^{-\frac{2\pi}{\beta} t_R},\vec{x}_{d-2}\}\ , \nonumber\\
&\{x^{-},x^+,\vec{x}_{d-2}\}_R=\{-re^{\frac{2\pi}{\beta} t_R},r e^{-\frac{2\pi}{\beta} t_R},\vec{x}_{d-2}\}\ 
\end{align}
to describe left and right Rindler wedges, respectively, implying $\sigma=e^{-\frac{2\pi}{\beta} t_R}/r$ and $\eta=r$.  

The Minkowski correlator \eqref{eq:correlator} now can  be viewed as a thermal correlator on Rindler space. In particular, we identify \cite{Maldacena:2015waa}
\be\label{otoc:Regge}
G(\eta,\sigma)=\frac{F\(t_R-\frac{i\beta}{4}\)}{F_d}\ .
\ee
Moreover, we are interested in the CFT Regge limit  which can be reached by taking 
\be\label{regge}
 \sigma\rightarrow 0 \ , \qquad \text{with } \qquad \eta=\text{fixed}>0
\ee
in the correlator $G(\eta,\sigma)$. In Rindler coordinate, this limit is equivalently described as $t_R\gg \beta$ of the OTOC $F\(t_R-\frac{i\beta}{4}\)$ with fixed $r=\eta$. It is easy to see that in the limit $t_R\gg \beta$, the factorization (\ref{eq:factorization}) follows trivially since the corresponding CFT correlator is determined by Euclidean OPE. So, the bounds from preceding sections apply to  the CFT Regge correlator (\ref{otoc:Regge}). 

We can now utilize the bounds from section \ref{sec:special} to constrain certain CFT Regge correlators. Consider a Regge correlator with the following Regge behavior for some range of $\sigma$:\footnote{We have defined $\sigma_*=\frac{1}{r}e^{-\frac{2\pi}{\beta} t_f}$, where $t_f$ is the cut-off in equation (\ref{eq:specialcase}).}
\be\label{con_anc}
G(\eta,\sigma)= 1+ i \sum_{L=2,4,6,\cdots}  \frac{c_L(\eta)}{ \sigma^{L-1}}\ , \qquad \sigma_*\le |\sigma|\ll \eta<1\ .
\ee
This Regge correlator is exactly equivalent to the OTOC (\ref{eq:specialcase}), once we identify (\ref{otoc:Regge}). From section \ref{sec:special}, we conclude that $c_L(\eta)$ coefficients must obey
\be\label{CFT:constraints}
c_L(\eta)>0\ , \qquad \frac{c_{L+2}(\eta) }{c_L(\eta)} < \sigma_*^{2}\ , \qquad c_{L+2}(\eta)^2 \le c_L(\eta) c_{L+4}(\eta)
\ee
for even $L\ge 2$. This agrees with bounds obtained in \cite{Kundu:2020gkz,Kundu:2021qpi} for interacting CFTs.

For CFTs that are dual to effective field theories in AdS with higher derivative four-point interactions, the Regge correlator has the exact form (\ref{con_anc}) \cite{JP,Chandorkar:2021viw,Kundu:2021qpi}. Note that odd spin contributions are not allowed in the expansion (\ref{con_anc}) because of symmetry.\footnote{More precisely, only operators with even spins can appear in the OPE of two identical operators.} Hence, these CFT Regge correlators saturate the bound (\ref{bound:intro}). In addition, the CFT constraints (\ref{CFT:constraints}) lead to bounds on various coupling constants of higher derivative four-point interactions of low energy effective field theories from UV consistency \cite{Kundu:2021qpi}.

\section{Conclusions}\label{sec:conclusions}
Remarkably, there is a universal bound \cite{Maldacena:2015waa} on the rate of growth of the OTOC (\ref{eq:otoc}) that follows directly from analyticity, positivity, and Schwarz reflection properties (\ref{halfstrip})-(\ref{eq:schwarz}) of the OTOC. However, the dispersion relation (\ref{OTOC:sumrule}) and the associated consistency conditions immediately imply that the OTOC must satisfy many more constraints beyond the MSS bound. Motivated by this observation, we performed a systematic analysis of the OTOC (\ref{eq:otoc}) starting from the basic properties (\ref{halfstrip})-(\ref{eq:schwarz}). We found that the complete set of bounds can be nicely organized in terms of the moment $\mu_J(t)$ of the OTOC. We proved that the moment  $\mu_J(t)$, as a function of integer $J\ge 0$, is positive, bounded, monotonically decreasing, and log-convex for any fixed $t\ge t_0$. The MSS bound can be obtained from this infinite set of conditions as the leading constraint. This provides an alternative derivation of the MSS bound that requires a weaker factorization assumption than what used in \cite{Maldacena:2015waa}. The subleading chaos bounds are also highly constraining, especially when the MSS bound is saturated. Still, it is not clear whether the bounds obtained in this paper are optimal or there are additional constraints on the OTOC.

One unsatisfactory  aspect of the formulation of chaos bounds in terms of the moment $\mu_J(t)$ is that the bounds (\ref{intro:positive})-(\ref{intro:log}) appear to be less transparent. Unlike the MSS bound, the full physical meaning of these new bounds is not so clear. Nevertheless, one can think of (\ref{intro:positive})-(\ref{intro:log}) as a generalization of the MSS bound that also constrains higher time-derivatives of $F(t)$. After all, inequalities (\ref{intro:positive})-(\ref{intro:log}) contain the following conditions: 
\be\label{conc:bound}
 \( \frac{2\pi}{\beta}- \p_t\) \( \frac{6\pi}{\beta}- \p_t\)\cdots \( \frac{2\pi(2N-1)}{\beta}- \p_t\)\(F_d-F(t)\)\ge 0\ ,  \qquad  t\gg  t_0
\ee
for all integers $N\ge 1$. For example, the bound (\ref{bound:intro}) can also be derived from the above chaos bounds with $N=1$ and $2$. However, we want to emphasize that the above bounds (\ref{conc:bound}), in general, are weaker than the bounds (\ref{intro:positive})-(\ref{intro:log}). The eigenstate thermalization hypothesis might be a good guide in further understanding what these additional constraints physically mean.

From the perspective of the bounds of this paper, the idea of maximal chaos appears to be rather arbitrary since it saturates only one of an infinite set of constraints. In fact, the contribution  $F_d-F(t)\sim \exp(\frac{2\pi}{\beta} t)$ alone, in any time duration, is inconsistent with the chaos bounds (\ref{intro:positive})-(\ref{intro:log}). In other words, the MSS bound (\ref{intro:MSS}) can never be exactly saturated.  At this point, one may wish to consider saturations of higher derivative cousins (\ref{conc:bound}) of the MSS bound. However, it is easy to show that an OTOC exactly saturating (\ref{conc:bound}) for any finite $N$ is also inconsistent with the bound (\ref{intro:positive}), since associated moments $\mu_{J\ge N}(t)=0$. On the other hand, the constraint  (\ref{conc:bound}) can be consistently saturated for $N=\infty$. In particular, this constraint is saturated exactly by the OTOC (\ref{eq:specialcase}), which we have analyzed in section \ref{sec:special}. So, the OTOC (\ref{eq:specialcase})  has an interpretation as a natural extension of the maximally chaotic OTOC. From this perspective, it is rather satisfying that this special OTOC naturally appears when we study Regge correlators in CFTs that are dual to effective field theories in AdS.

The chaos bounds obtained in this paper will be important when we study general analytic completions of maximal chaos. We will explore this in a separate paper \cite{Kundu:2021mex}. We believe that techniques utilized in this paper to derive bounds will have other general applications. For example, these tools can be useful for analyzing chaos inside  the butterfly cone, which is generally characterized by a velocity-dependent Lyapunov exponent \cite{Xu:2018xfz,Khemani:2018sdn}. It is possible that there are additional constraints beyond the bound obtained in \cite{Mezei:2019dfv}. Perhaps, a similar analysis can also be performed for the spectral form factor \cite{Cotler:2016fpe}, though we will have to leave this problem for the future.

%%%%%%%%%%%%%%%%%%%%%%%%%%%%%
\section*{Acknowledgments}

It is my pleasure to thank Diptarka Das, Thomas Hartman, Jared Kaplan,  Arnab Kundu, Wei Li, Eric Perlmutter, and Douglas Stanford  for helpful discussions and commenting on a draft. I was supported in part by the Simons Collaboration Grant on the Non-Perturbative Bootstrap.

\begin{appendix}

\section{Moments with Non-Integer $J$}\label{app:noninteger}
There is no positivity condition on $\mu_J\(t\)$ when $J$ is not a non-negative integer. However, that does not mean there are no constraints. Absolute values of moments for any real $J$ are bounded from above. To see that let us first use the analyticity property to write 
\begin{align}
\mu_J(t)=2e^{\frac{4\pi J t}{\beta}}e^{i \pi J}\int_{t}^\infty dt' e^{-\frac{2\pi  (2J+1)}{\beta}t'}\mbox{Re}\left[e^{-i \pi J}\(F_d- F(t'+i\beta/4)\)\right]
\end{align}
for any real $J>-\frac{1}{2}$. Note that $\mu_J(t)$ is completely imaginary when $J$ is a positive half-integer. Moreover, we also find that for any $J>-\frac{1}{2}$
\begin{align}\label{eq:allbound}
|\mu_J(t)|&\le2e^{\frac{4\pi J t}{\beta}}\int_{t}^\infty dt' e^{-\frac{2\pi  (2J+1)}{\beta}t'}|F_d- F(t'+i\beta/4)|\nonumber\\
&<\frac{2\beta F_d}{\pi(2J+1)}e^{-\frac{2\pi}{\beta}t}\ .
\end{align} 

Finally, let us make few comments for moments with negative $J\le -\frac{1}{2}$. For any real $J$, we can generalize (\ref{eq:rate}) to obtain 
\be
e^{\frac{4\pi J t}{\beta}}\ \frac{d}{dt} \(e^{-\frac{4\pi J t}{\beta}}\mu_J(t)\)=-2 e^{-\frac{2\pi  }{\beta}t+i \pi J}\mbox{Re}\left[e^{-i \pi J}\(F_d- F(t'+i\beta/4)\)\right]\ .
\ee

\section{An Expansion for  $\mbox{Im}\ t =\pm \frac{\beta}{4}$}\label{app:boundary}
As we mentioned before, the expansion (\ref{eq:master}) is not valid for $\mbox{Im}\ t =\pm \frac{\beta}{4}$. However, we show next that the expansion (\ref{eq:master}) correctly reproduces the imaginary part of the OTOC even at $\mbox{Im}\ t =\pm \frac{\beta}{4}$. 

We are now focus on the special case $|\mbox{Im}\ t |= \frac{\beta}{4}$:
\be
\oint_\gamma dt' \ e^{\frac{2\pi }{\beta}(t'-t)} \frac{F(t')-F_d}{1-e^{\frac{4\pi }{\beta}(t'-t)}}=0 \ ,
\ee
where, the contour $\gamma$ is now designed to avoid the singularities at $t'=\mbox{Re}\ t \pm i \frac{\beta}{4}$ by including small semicircles around them. This contour integral leads to 
\begin{align}\label{eq:im_boundary}
\mbox{Im}\ F\(t+i\frac{\beta}{4}\)=-\frac{2}{\beta} e^{\frac{2\pi }{\beta}t} \sum_{J=0,\pm 1,\pm 2, \cdots} \mu_J( t) 
\end{align}
for real $t$. Note that this agrees with the expansion (\ref{eq:master}). 

A similar expression can be obtained for the real part as well by using equation (\ref{eq:rate}). In particular, for real $t$ we obtain 
\be\label{app:real}
\mbox{Re}\ F\(t+i\frac{\beta}{4}\)=F_d+\frac{1}{2}  e^{\frac{2\pi  }{\beta}t} \mu_0'(t)\ .
\ee

\end{appendix}

%%%%%%%%%%%%%%%%%%%%%%%%%%%%%%%%%%%%%%%%%%%

\end{spacing}

\bibliographystyle{utphys} 
\bibliography{chaos}

\end{document}